\documentclass[manuscript,screen,authorversion,nonacm]{acmart}
\AtBeginDocument{%
  \providecommand\BibTeX{{%
    \normalfont B\kern-0.5em{\scshape i\kern-0.25em b}\kern-0.8em\TeX}}}

\copyrightyear{2023} 
\acmYear{2023} 
\setcopyright{rightsretained} 
\acmConference[CHI EA '23]{Extended Abstracts of the 2023 CHI Conference on Human Factors in Computing Systems}{April 23--28, 2023}{Hamburg, Germany}
\acmBooktitle{Extended Abstracts of the 2023 CHI Conference on Human Factors in Computing Systems (CHI EA '23), April 23--28, 2023, Hamburg, Germany}\acmDOI{10.1145/3544549.3585713}
\acmISBN{978-1-4503-9422-2/23/04}

\begin{document}

\title{Trust and Reliance in Consensus-Based Explanations from an Anti-Misinformation Agent}


\author{Takane Ueno}
\email{ueno.t.ao@m.titech.ac.jp}
\orcid{1234-5678-9012}
\affiliation{
  \institution{Tokyo Institute of Technology}
  \streetaddress{2 Chome-12-1 Ookayama}
  \city{Meguro City}
  \state{Tokyo}
  \country{Japan}
  \postcode{152-8550}
}

\author{Yeongdae Kim}
\email{kim.y.ah@m.titech.ac.jp}
\orcid{0000-0001-5346-0041}
\affiliation{
  \institution{Tokyo Institute of Technology}
  \streetaddress{2 Chome-12-1 Ookayama}
  \city{Meguro City}
  \state{Tokyo}
  \country{Japan}
  \postcode{152-8550}
}

\author{Hiroki Oura}
\email{houra@rs.tus.ac.jp}
\orcid{0000-0001-5111-7864}
\affiliation{
 \institution{Tokyo University of Science}
 \streetaddress{1 Chome-3 Kagurazaka}
 \city{Shinjuku City}
 \state{Tokyo}
 \country{Japan}
 \postcode{162-8601}}

\author{Katie Seaborn}
\email{seaborn.k.aa@m.titech.ac.jp}
\orcid{0000-0002-7812-9096}
\affiliation{
  \institution{Tokyo Institute of Technology}
  \streetaddress{2 Chome-12-1 Ookayama}
  \city{Meguro City}
  \state{Tokyo}
  \country{Japan}
  \postcode{152-8550}
}

\renewcommand{\shortauthors}{Ueno et al.}

\begin{abstract}
 The illusion of consensus 
 occurs when people believe there is consensus across multiple sources, but the sources are the same and thus there is no "true" consensus. We explore this phenomenon in the context of an AI-based intelligent agent designed to augment metacognition on social media. Misinformation, especially on platforms like Twitter, is a global problem for which there is currently no good solution. 
 As an explainable AI (XAI) system, the agent provides explanations for its decisions on the misinformed nature of social media content. In this late-breaking study, we explored the roles of trust (attitude) and reliance (behaviour) as key elements of XAI user experience (UX) and whether these influenced the illusion of consensus.
Findings show no effect of trust, but an effect of reliance on consensus-based explanations. This work may guide the design of anti-misinformation systems that use XAI, especially the user-centred design of explanations.

\end{abstract}





\maketitle

\section*{Citation}

\noindent
Takane Ueno, Yeongdae Kim, Hiroki Oura, and Katie Seaborn. 2023. Trust and Reliance in Consensus-Based Explanations from an Anti-Misinformation Agent. In \textit{Extended Abstracts of the 2023 CHI Conference on Human Factors in Computing Systems (CHI EA '23)}. Association for Computing Machinery, New York, NY, USA, Article 296, 1–7. \url{https://doi.org/10.1145/3544549.3585713} \\

\noindent
The final publication is available via ACM at \url{https://dl.acm.org/doi/10.1145/3544549.3585713}.

\section{Introduction}
AI is increasingly being taken up for a variety of purposes, thanks to complex algorithms and the availability of big data. At the same time, a user-centred problem with AI is coming into sharper relief: that the algorithms making up AI systems are usually black boxes that make decisions in ways not transparent to people. Even when they are transparent, people are often unable to understand because the output is too complex. Indeed, prediction accuracy and explainability in AI are generally trade-offs \cite{rai_explainable_2020}. To address these issues, the notion of explainable AI (XAI) has gained momentum in recent years. The purpose of XAI, a term coined by DARPA researchers in 2017, is to make AI decision-making and other fundamental behavior more understandable to people by providing a human-parsable explanations \cite{gunning_darpas_2021}. 

Many of us, members of the general public, are starting to use these seems (or desire to). For example, many people have become concerned with the level of misinformation on social media and desire intelligent fact-checking tools \cite{urakami_finding_2022}, notably on Twitter, and perhaps especially now after its acquisition by Elon Musk and subsequent revocation of fail-safe measures \cite{the_associated_press_twitter_2022}. Still, few XAI initiatives and specifically the forms of the explanations provided are designed with general non-professional end-users in mind, i.e., the lay public. A recent survey found that machine learning (ML) model descriptions are primarily used by ML engineers to debug models during the development phase \cite{bhatt_explainable_2020}. Furthermore, most work appears to rely on the researchers' own intuitions, i.e., expertise, of what constitutes a good explanation \cite{miller_explanation_2019}.

A pressing question is what forms of explanations are effective at enabling lay end-users to trust XAI systems \cite{brennen_what_2020,weitz_you_2019, weitz_let_2021, ueno_trust_2022}. In social media platforms, lay people make judgments and decisions based on evaluating and integrating reports from multiple informants \cite{urakami_finding_2022, connor_desai_getting_2022}. One factor that affects lay people's confidence in such reports is the degree of consensus across related reports \cite{connor_desai_getting_2022}. In social media, people may desire explanations that rely on consensus-based data sources. Placing these sources within explanations provided by anti-misinformation XAI may increase users' confidence in the system's fact-checking and possibly reliance on it. Yet, this poses a new problem: an \emph{illusion of consensus} effect \cite{connor_desai_getting_2022,yousif_illusion_2019}, where people are unable to distinguish ``true'' consensus, i.e., different informants relying on different sources but drawing the same conclusion, and ``false'' consensus, i.e., different informants relying on the \emph{same} source. 

This raises three questions at the intersection of user-centred design and research methods. First, are people able to distinguish true and false consensus, i.e., not fall prey to an illusion of consensus, when explanations are provided by a user-centred XAI system? Second, how do we distinguish those who are simply skeptical of AI? And third, if the illusion occurs, can it be decreased by emphasizing the independence of sources via the design of the explanations, as suggested by work outside of social media \cite{connor_desai_getting_2022}? To the best of our knowledge, this has not yet been explored.
Therefore, we asked: \emph{How does consensus relate to user trust and reliance on the use of an intelligent anti-misinformation XAI agent?} Specifically: \emph{(RQ1) If the agent provides a consensus-based explanation about its fact-checking, does this lead to increased trust (attitude) and/or reliance on the agent (behavior)?} We also asked: \emph{(RQ2): If an illusion of consensus appears, what effect, if any, does the consensus-based explanation have on trust and/or reliance?}.
To this end, we conducted a comparative evaluation of a prototypical XAI agent within a live Twitter environment that provided explanations to lay Twitter users about its fact-checking decisions.  
The main contributions of this work are: (i) initial empirical attitudinal, behavioural, and user experience (UX) evidence of a relationship between reliance, but not trust, on fact-checking services provided by a consensus-based XAI agent and subsequently (ii) evidence of an illusion of consensus effect. This work highlights the importance of consensus and its presentation in XAI systems using the case study of misinformation on Twitter.

\section{Background}
\subsection{Trust and Reliance in AI}
Trust in automation is defined as ``the attitude that an agent will help achieve an individual’s goals in a situation characterized by uncertainty and vulnerability'' \cite[p.~54]{lee_trust_2004}.
This definition is based on a strong foundation of empirical research and frequently referenced in the context of AI \cite{ueno_trust_2022}. 
Lee and See \cite{lee_trust_2004} define trust as an \emph{attitude} and distinguish it from \emph{trust} as a behavior.
Similarly, Hoff and Bashir's \cite{hoff_trust_2015} model of trust in automation makes a clear distinction between trust as an attitude and behavior, characterizing trust as a factor that mediates automation performance and user behavior, or \emph{reliance}. Most recently, Papenmeier, Kern, Englebienne and Seifert \cite{papenmeier_its_2022} reported discrepancies between self-reported trust and trust as a behavior, indicating that it is important to clearly distinguish between attitude and behavior. In this work, we follow suit by operationalizing trust as an attitude and reliance as a behavior.

Still, these concepts were developed for automation. Indeed, trust and reliance have not been well distinguished in research on AI-based systems and have been grouped together under the term trust \cite{ueno_trust_2022}. Trust is typically measured in isolation as a subjective attitude through questionnaires and interviews \cite{kunkel_let_2019, hoggenmuller_context-based_2021, weitz_let_2021}. In some studies, trust has been approached as \emph{dependent} behaviors or biological responses  \cite{waytz_mind_2014, okamura_calibrating_2020}. 
Other research has considered the roles of automation performance and dependence as mediators of trust \cite{wiczorek_is_2010, chancey_role_2013, merritt_affective_2011}. The focus has been on the relationship between automation reliability, dependence, and trust, and the results of these studies are somewhat contradictory. Hussein, Elsawah and Abbass \cite{hussein_trust_2020} reexamined this literature and developed experimental guidelines to reduce errors. They analyzed the role of trust mediation on perceived reliability of and dependence on a target sensing system in a flight task. In order to clearly distinguish between trust and reliance, we applied their guidelines to explore the relationship between these factors in the context of consensus-based explanations provided by XAI.



\subsection{Appraising Consensus Across Reports and Sources}
In daily life, we rely on consensus when evaluating and integrating various pieces of information to make decisions \cite{connor_desai_getting_2022}.
However, information is not always independent; separate pieces of information may use the same source/s. For example, over 80\% of climate change denial blogs relied on a single primary source \cite{harvey_internet_2018}. Reliance by multiple independent informants on a single source of data is called a \emph{false consensus} and can influence the formation of accurate beliefs. Yousif, Aboody and Keil \cite{yousif_illusion_2019} investigated perceptions false consensus. Subjects were were assigned to one of a true consensus condition in which they read four positive sentences with different primary sources and one sentence with a negative primary source, a false consensus condition in which they read four positive sentences with a single primary source and one sentence with a negative primary source, a false consensus condition in which they read a positive sentence and a negative, and a baseline condition in which they read one sentence each, and after reading each sentence, they were asked how much they agreed with the assertion. As a result, they discovered an \emph{illusion of consensus}, in which subjects gave similar agreement ratings to presentations of true and false consensus.
Connor Desai, Xie, and Hayes \cite{connor_desai_getting_2022} investigated this consensus illusion, believing that its creation was due to people's perception of the independence of information sources. They followed the same experimental procedure as Yousif et al. \cite{yousif_illusion_2019} but also highlighted single sources of information with the same color, emphasizing the relationship between each source. As a result, true consensus, with its emphasis on independence, received greater agreement than false consensus. They further investigated this in the context of an election poll on Twitter and showed that people assigned more epistemic weight to true consensus than to false consensus when the relationship between sources was made transparent.
Consensus thus appears to affect our level of agreement with an opinion. But do these findings translate to consensus-based explanations provided by XAI? Depending on the type of consensus provided, people may assign different levels of agreement to XAI explanations, thus mediating their trust and reliance on the XAI. Also, if these explanations are transparent about the relationship between data sources, the illusion of consensus may not occur, even when the agent provides the consensus. We follow Hussein et al. \cite{hussein_trust_2020}’s experimental design by extending the theory of Yousif et al. \cite{yousif_illusion_2019} and Connor Desai et al. \cite{connor_desai_getting_2022} to explore the XAI context and these possible effects in this work.


\section{Theoretical Framework}
\begin{figure}[h]
    \centering
    \includegraphics[width=.3\textwidth]{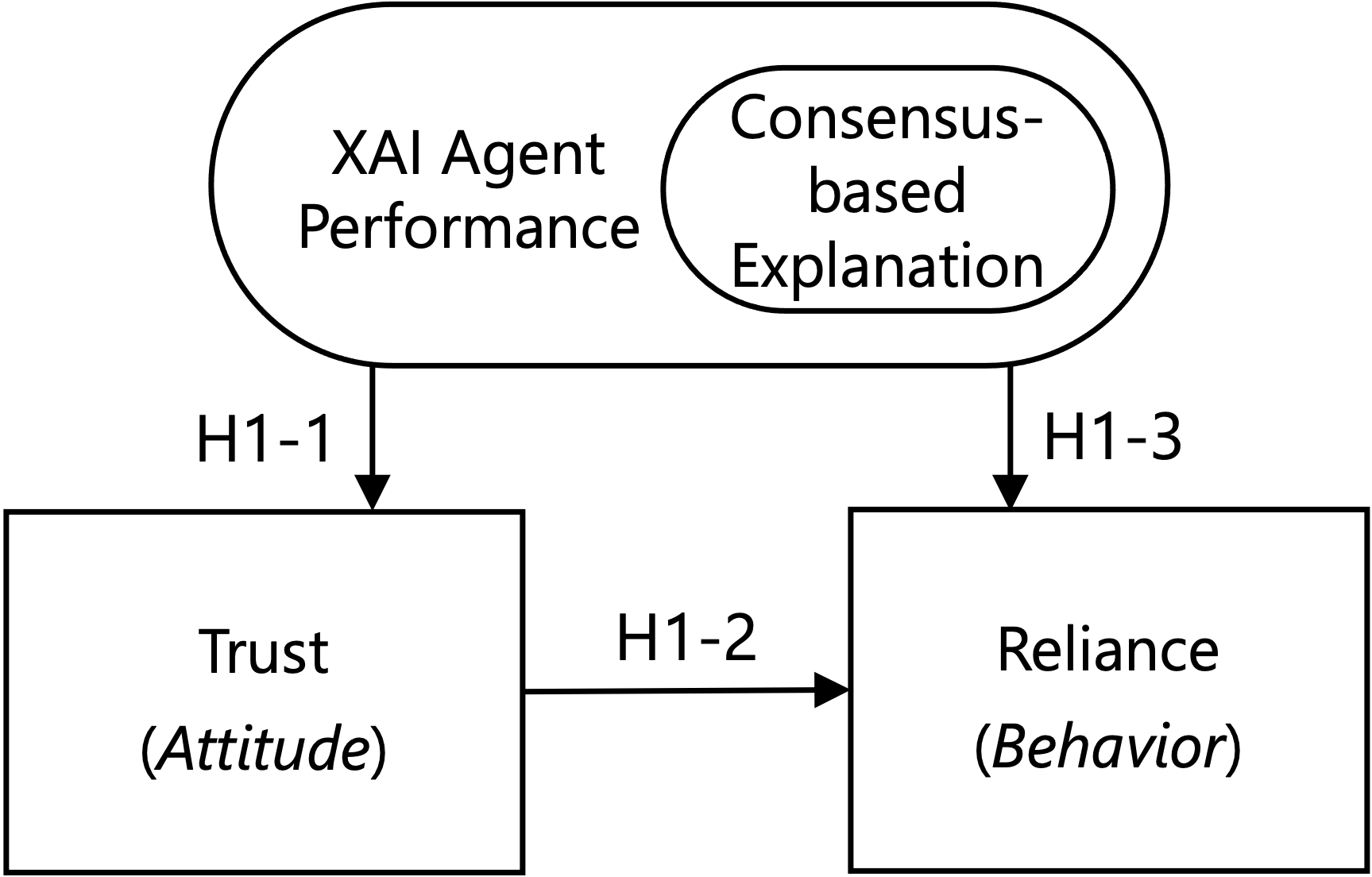}
    \Description[Relationship diagram of the three elements]{XAI Agent Performance, Trust, and Reliance are interrelated. XAI Agent Performance includes Consensus-based Explanations, which affect Trust and Reliance, respectively. Also, Trust affects Reliance. These relationships are expressed as Hypotheses 1, 2, and 3, respectively.}
    \caption{Our theoretical framework, based on Hoff and Bashir's model of trust for automation \cite{hoff_trust_2015}. Consensus-based explanations provided by the XAI agent influence reliance (behavior), which is mediated by trust (attitude) in the agent's performance.}
    \label{fig:TrustModel}
\end{figure}

Given the lack of established models of trust in AI and particularly XAI \cite{ueno_trust_2022}, we used Hoff and Bashir's model for automation \cite{hoff_trust_2015, hussein_trust_2020}. Our instantiation of the model for our XAI agent and construction of hypotheses is in Figure \ref{fig:TrustModel}.

Agreement with information when there is true consensus tends to be higher than when there is false or no consensus \cite{connor_desai_getting_2022, yousif_illusion_2019}. Still, if the trust model for automation \cite{hoff_trust_2015, hussein_trust_2020} applies to XAI systems, then we would expect reliance behavior to be mediated by trust. We thus hypothesize: \emph{(H1-1) ``True'' consensus-based explanations from agent will increase user trust in agent compared to ``false'' consensus and ``no'' consensus.} This effect may also apply to the agent-provided explanations and be reflected in user reliance on the agent, leading to this hypothesis: \emph{(H1-2) User trust in the agent increases user reliance on the agent.} Subsequently, if the results hold true for ``true'' consensus, and the theorized relationship between trust and reliance exits for XAI systems, then we can also hypothesize: \emph{(H1-3) ``True'' consensus-based explanations from the agent will increase user reliance on the agent compared to ``false'' consensus and ``no'' consensus.}

Following previous work \cite{connor_desai_getting_2022}, our XAI agent explicitly \emph{labels} sources of information and the data used by these sources to clarify the relationship between individual sources and data. In other words, these labels of sources and data should clearly indicate to users that each source is independent in a true consensus conditions and not independent in a false consensus conditions. Emphasizing the independence between data across sources reduces the illusion of consensus \cite{connor_desai_getting_2022}. Thus, the design of the XAI agent's explanations should prevent the illusion of consensus from occurring. We hypothesize: \emph{(H2) The illusion of consensus will not appear when the source information about the data is explicit.}

\section{Method}
We conducted a within-subjects experiment based on Hussein et al. \cite{hussein_trust_2020}. We used an intelligent XAI agent designed to support metacognitive behaviors in the face of misinformation on Twitter, specifically related to the COVID-19 pandemic. Our protocol was registered in advance of data collection on July 7\textsuperscript{th}, 2022\footnote{\url{https://osf.io/s7wqe}}. We obtained ethics from our IRB.




\subsection{Participants}
A total of 35 participants (22 men, 13 women, none who identified as another gender) who were fluent in Japanese and used Twitter were recruited. The sample size was determined based on the previous study by Hussein et al. \cite{hussein_trust_2020}. Participants were recruited from Jikken-baito, a Japanese experiment recruiting website through multiple social media platforms and connection between researchers\footnote{\url{https://www.jikken-baito.com}} or directly by the authors.

\subsection{System Design}
\begin{figure}[h]
    \centering
    \includegraphics[width = .70\textwidth]{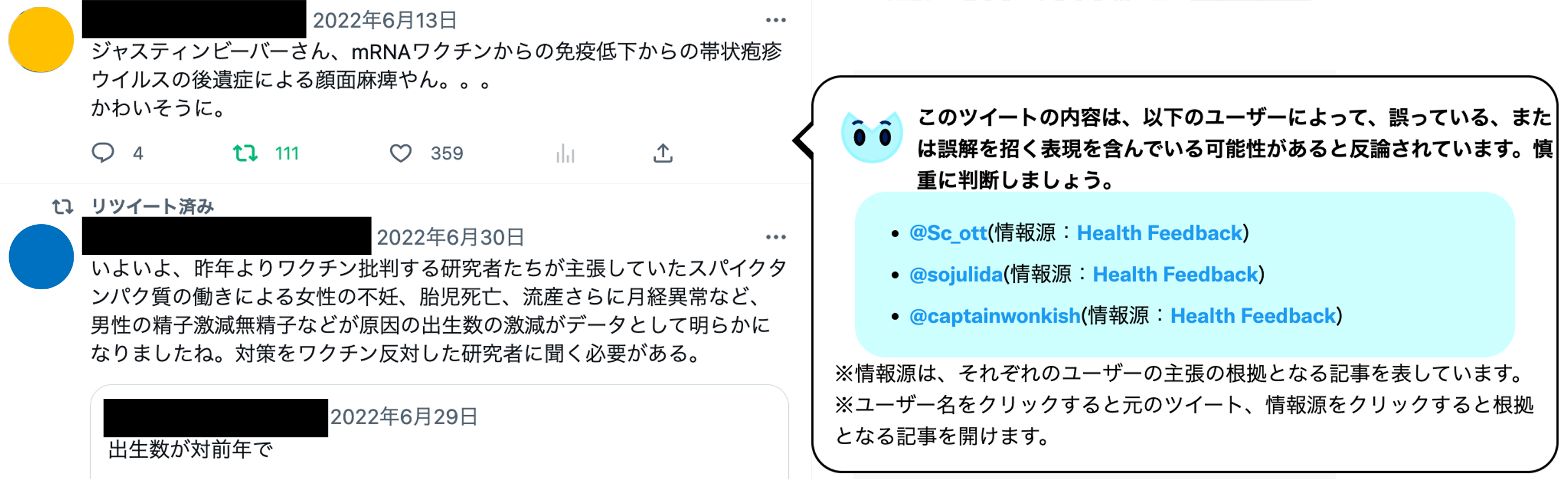}
    \Description[Agent example]{The agent's banner appears on the right side of the Japanese Twitter timeline.
The agent writes in Japanese, "このツイートの内容は、以下のユーザーによって、誤っている、または誤解を招く表現を含んでいる可能性があると反論されています。慎重に判断しましょう。" 
Additionally, the agent provides the IDs of three different other users and one identical source they cite.}
    \caption{The agent creates a banner on the right side of the misinformed tweet to display an explanation with references to tweets and other information sources and data. These were not restricted to Twitter alone and included external sources. Here, a ``false'' consensus of three different tweets referring to the same source is shown.}
    \label{fig:SystemDesign}
\end{figure}

We used a novel Twitter-based intelligent XAI agent called Elemi \cite{kim_exoskeleton_2023}. The agent, which requires curated content, simulates fact-checking within tweet content, providing links to other tweets and sources. If a tweet contains misinformation, the agent adds a banner to the top-right side of the tweet containing an explanation with tweets and their data sources as references for why the agent regards the tweet as misinformed. As a simulated agent that uses curated content, its accuracy is 100\%. Tweets and sources were manually collected and verified by the authors. There were three consensus schemes: \emph{True} (three tweets referring to three different sources), \emph{False} (three different tweets referring to the same source), and \emph{None} (only one tweet). Refer to Figure \ref{fig:SystemDesign} for an illustration of the agent in action.


\subsection{Stimuli}
Participants viewed a controlled Twitter timeline with tweets sourced from the COVID-19 hashtag in September 2022: 12 factual and 12 misinformed. The factualness of the tweets was verified by the first author based on at least two different sources. Since we used live Twitter, it was possible for participants to have come across these tweets before. Factual and false tweets were randomly ordered into pairs. No other tweets were in the timeline.

\subsection{Procedure}
All participants gave informed consent.
The experiment was divided into two sessions. In Session 1, participants verbally answered how correct they thought each tweet was on a scale of 0-100 (0: completely wrong, 100: completely right) while viewing the timeline \emph{without} the agent. Then there was a 5-minute break.

In Session 2, participants carried out the same procedure.
However, this time false tweets were pointed out by the agent. 
The agent's three consensus conditions (``true'', ``false'', ``no'') were counterbalanced and changed every eight tweets to account for individual differences.
 Note that, due to the Musk acquisition of Twitter, some tweets which the agent serve as data sources unexpectedly became unavailable. To mitigate the impact on trust and reliance, participants who witnessed tweets being unavailable were asked to imagine the existence of other tweets similar to those provided by the agent. To accommodate the dynamic nature of trust \cite{hussein_trust_2020}, participants paused after reading two tweets (one factual and one false) and completed a questionnaire with the trust measures (4.5.1) on a separate tablet.


After the sessions, participants completed a post-experiment questionnaire that included demographics and open-ended questions: ``How did you feel overall about your experience with the agent (Session 2)? Why did you feel that way? Please be specific.'' ``What did you feel about the data that the agent used to identify tweets with potentially false content? Why did you feel that way? Please be specific.'' Participants were then thanked and compensated.

\subsection{Measures}
All measures were translated into Japanese by the authors and back-translated using DeepL, checked by those fluent in both languages. All references to ``the system'' in the instruments were changed to the agent's name.

\subsubsection{Trust}
We used the Trust in Automation scale \cite{jian_foundations_2000}, a 7-point Likert scale consisting of 12 items: 7 for
trust and 5 for distrust. Although developed for automation, it is also the most commonly used measure for AI \cite{ueno_trust_2022}. Trust and distrust can exist simultaneously and are different concepts \cite{lewicki_trust_1998, sitkin_explaining_1993}. In our case, we measured trust multiple times (refer to 4.4). In consideration of participant time and workload, only the seven items related to trust were used.

\subsubsection{Reliance}
Reliance was measured using Weights of Advice (WOA) \cite{harvey_taking_1997}. While WOA has been used in AI and XAI research as a measure of trust \cite{panigutti_understanding_2022, logg_algorithm_2019, poursabzi-sangdeh_manipulating_2021}, we used it as a measure of reliance because it is an objective indicator of behavior rather than subjective and attitudinal. 
WOA 
quantifies the extent to which participants change their ratings as a result of an informant's advice: $WOA_{ij} = (F_{ij}-I_{ij})/(A_{ij}-I{ij})$, where I, F, and A denote the initial estimate, the final estimate, and the advisor's advice for some participant i on some trial j, respectively. A WOA of 1 indicates adoption of the advice, 0 indicates maintenance of the initial estimate, and between 0 and 1 indicates that the advice is partially discounted. Notably, a WOA of 0.5 indicates equal weighting of one's own estimate and the advisor's advice. The values of Session 1 and Session 2 were assigned to I and F, respectively. The agent gave advice only on tweets that contained false content, so the WOA was computed only for tweets containing false content and was fixed at A = 0. The value of WOA was truncated to 0 for values less than 0 and to 1 for values greater than 1, following previous studies \cite{gino_effects_2007, gino_we_2008, logg_algorithm_2019}. Note that some previous studies used absolute values when calculating WOA measurements. For robustness, we also used the absolute value approach, but the nature and significance of the results remain the same.


\subsection{Data Analysis}
Data were measured by subject for each condition. 
One person's data was excluded because they rated all measures at 50\%, indicating an inability to make judgments about correctness. Data in columns with initial estimate (I) equal to the advice (A) were excluded 
according to previous work \cite{gino_effects_2007, gino_we_2008}. In the end, 289/315 points of data were analyzed. 
We fitted random intercept models; all models contain subject as a random effect and consensus (and possibly trust) as fixed effects. We also averaged the results per consensus condition and ran one-way repeated measures ANOVAs. 

An applied thematic analysis \cite{guest_applied_2012} was conducted on the open-ended responses to explore factors that may influence trust and reliance and compare awareness of the consensus and number of sources. A lead rater developed the initial themes, and then two raters coded all data separately. Inter-rater reliability was assessed by Cohen's kappa \cite{landis_measurement_1977} with 0.7+ as the criterion for agreement. Themes that did not meet this criterion were modified, merged, or discarded and repeated until the kappa exceeded 0.7. For coding that did not match, disagreements were resolved by discussion.

\section{Results and Discussion}
\begin{figure}[h]
    \centering
    \includegraphics[width=.46\textwidth]{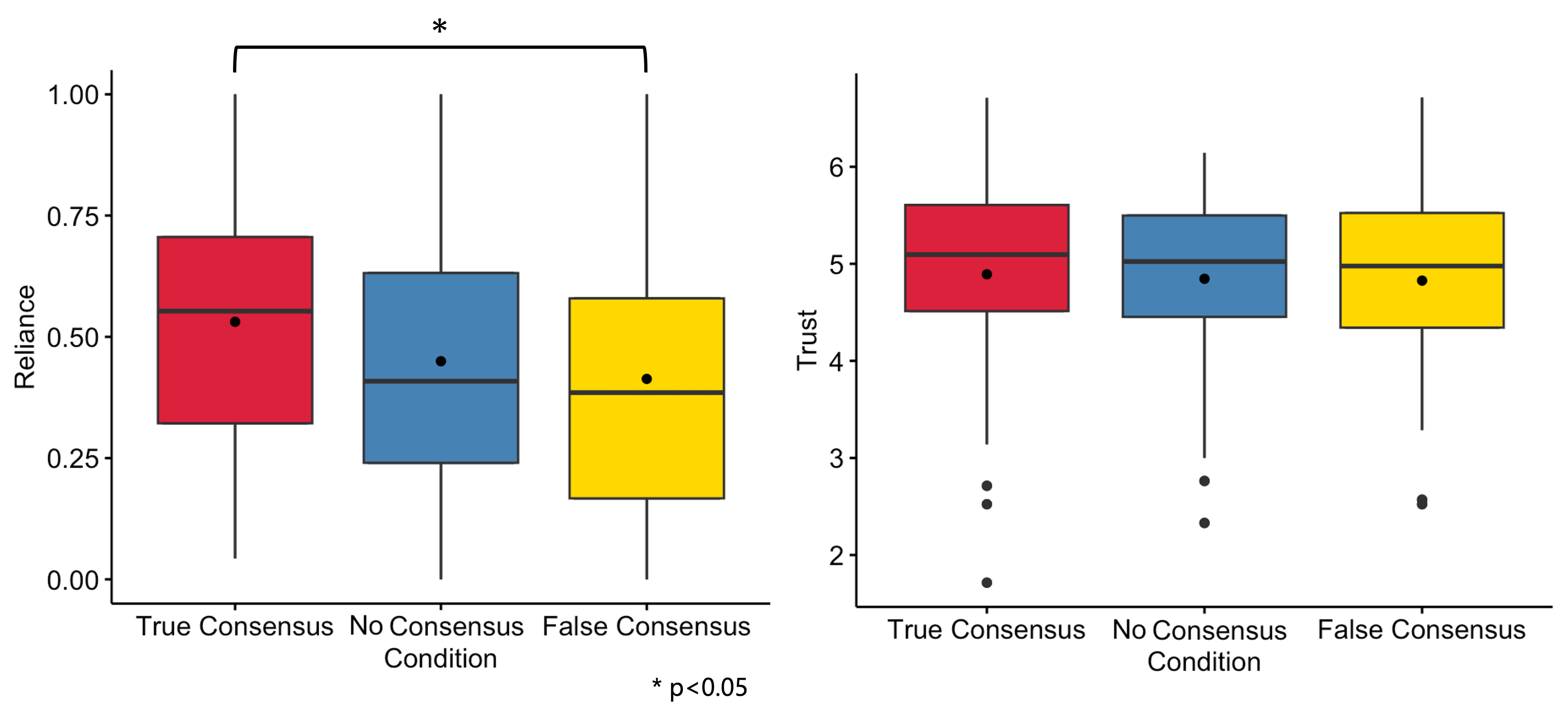}
    \Description[Six box-and-whisker plots]{Box-and-whisker plots of Reliance and Trust for each of the three conditions: true consensus, false consensus, and no consensus. True consensus and false consensus only for reliance show significant differences in ANOVA.}
    \caption{The box-and-whisker diagram shows the results of the ANOVAs. The boxes indicate the interquartile ranges and the horizontal lines are the medians. The circles inside the box represent the mean.}
    \label{fig:boxplot}
\end{figure}

Figure \ref{fig:boxplot} and Table \ref{tab:Model} summarize the results of the statistical analyses. Table \ref{tab:thematic analysis} shows the thematic framework.

We begin with the linear mixed model. The Cronbach's alpha for trust was $\alpha$ = 0.90 and the intraclass correlation coefficient (ICC) for trust was 0.65.
In the trust model $M_0$, consensus is incorporated as a dummy variable based on the true condition because of the categorical three conditions. Neither coefficient was statistically significant (\emph{p} > .05)
 The ICC for reliance was 0.34. 
 In the reliance model $M_0$, consensus is incorporated, with $\beta_1$ not statistically significant (\emph{p} > .05) but $\beta_2$ = -0.33, 95\% confidence interval (CI) = [-0.56, -0.10], which was statistically significant (\emph{p} = .006). Reliance model $M_1$ was added trust to and was statistically significant with only $\beta_2$  = -0.33, \emph{p} = (.006), 95\% CI = [-0.56, -0.093].

 \begin{table*}
   \caption{Results 
   for the random intercept model. $\beta_1$-$\beta_3$ represent the coefficient of the no consensus dummy variable on true consensus, the coefficient of the false consensus dummy variable on true consensus, and the coefficient of trust, respectively. The marginal and conditional coefficients of determination $R^2_m$ and $R^2_c$ and the Akaike Information Criterion (AIC) were computed. $^*$: \emph{p} < .05}
  \label{tab:Model}
  \begin{tabular}{cccccccccc}
    \toprule
     Response&Model&$\beta_1$&$\beta_2$&$\beta_3$ &W-S & B-S& $R^2_m$ & $R^2_c$ &AIC\\
     Variables& & & & & Variance & Variance& &  &\\
     
     \midrule
     Trust & $M_0$ & -0.05 &-0.05 &- & 0.36 &0.66& 0.001 &0.65 & 638.6\\
     &  & [-0.22, 0.12]  & [-0.23, 0.12]  & & & & & & \\
     \hline
     Reliance & $M_0$ & -0.22 & $-0.33^*$ &- & 0.66 & 0.35 & 0.02 & 0.36 & 775.1\\
     &  & [-0.45, 0.01]  & [-0.56: -0.10]  & & & & & & \\
     \cline{2-10}
     & $M_1$ & -0.22 & $-0.33^*$ &0.023 & 0.67 & 0.36 & 0.02 & 0.36 & 780.6\\
     &  & [-0.45, 0.01]  & [-0.56, -0.09]  & [-0.11, 0.16]& & & & & \\

  \bottomrule
 \end{tabular}
 \end{table*}

 An ANOVA indicated a statistically significant difference across consensus conditions, \emph{F}(2, 66) = 4.46, \emph{p} = .015. A post-hoc analysis with a Bonferroni correction revealed that the pairwise differences between false and true were statistically significantly different (\emph{p} = .032). Other pairwise differences were not statistically significantly different (\emph{p} > .05). Also, trust was not statistically significant under different consensus conditions, \emph{F}(2, 66) = .10, \emph{p} = .90.

 Ten qualitative sub-themes were identified and further classified into two major themes (Table~\ref{tab:thematic analysis}). The theme "The agent and its algorithm" includes references related to trust, dependence, and reliability of the agent itself and its algorithm. The theme "Tweets/sources provided by the agent" refers to the tweets provided by the agent in its explanations, and references related to trust, dependence, and reliability on the fact-checking articles as sources.

The results partially support (H1-2) but do not support (H1-1) and (H1-3) with respect to (RQ1). 
True consensus significantly increased reliance on the agent compared to false consensus, but did not do better than no consensus. Also, consensus-based explanations did not affect trust in the agent. The agent consistently received high trust scores (\~5/7), which did not mediate the relationship between consensus-based explanations and reliance. In short, trust (attitude) towards the agent was not affected by consensus-based explanations. Still, subjects recognized the difference between true and false consensus and relied more on true consensus-based explanations (behavior). Consensus or number of sources was an important factor.
Even so, most subjects stated that the function was useful, which may be a factor that sustained such a high level of trust. Trust does not necessarily manifest itself in reliance, and reliance is not necessarily evidence of trust \cite{miller_explanation_2019}. Our results reiterate the importance of separating trust and reliance in XAI agents.
 
 \begin{table*}
 \caption{Thematic analysis framework. Numbers in parentheses are frequencies for Q1 and Q2, respectively. 
 }
\label{tab:thematic analysis}
  \begin{tabular}{lllll}
    \toprule
     Theme & Sub-theme & Definition & Examples \\
     
     \midrule
     The agent & Usefulness of  & The agent's ``fact-checking'' and sources &``It was useful to know which   \\
     and its & the agent  & were useful and effective, or not. & tweets might include false and   \\
     algorithm & (29, 5) &  & the tweets.'' 
    \\
     
     \cline{2-4}
     & Concerns about & Fear of or actual over\text{-}reliance and  & ``The agent almost made me  \\
     &  misuse of the  & misuse of the agent. &  decide that the information was \\
     &  agent (10, 2) &  &  false without checking the link.'' \\
     
     \cline{2-4}
     & Consistent with  & Agent and subject agreed on factualness & ``My own feeling matched the    \\
     & the agent (4, 2)  & of the content, or not. &  agent's many times, ''\\
       
     \cline{2-4}

     & Time lapse & Subjects' trust and reliance on the agent  & ``I was skeptical in the beginning,   \\
     & (3, 2) & changed over time. &  but I trusted it in the end …''\\

     \cline{2-4} 
     
     & Questions about  & Subjects wanted to understand the  & ``I wondered how they find users \\
     & the agent's  & agent's algorithm and the criteria by  & who have opposing views.'' \\
     & algorithm (2, 1)  & which it made decisions. & \\

    \hline
    Tweets/ & Reliability of  & Feelings about the reliability of the 
    & ``The agent's data sources also  \\
    sources & content (3, 20) & tweets and sources provided by the & seemed like sites I wasn't sure I  \\
    provided & & agent. & could trust.''  \\
    
    \cline{2-4}
     by the & Number/ & Trust and reliance on the agent varied  & ``I felt that it would be more  \\
     agent & consensus of  & based on the number of sources and   &  credible if the sources were not all   \\
     & sources (5, 8) & type of consensus the agent provided. & the same.'' \\
     
    \cline{2-4}
     & Unavailable  & When the agent provided unavailable  & ``…I have a sneaking suspicion that \\
     & tweets (1, 7) & tweets, subjects felt a loss of trust  & the fact that it was erased may have \\
     &  & and reliance on the agent. & been false information in itself…''\\

     \cline{2-4}
     & Unfamiliar & Difficult to judge the agents' support  & ``I was not familiar with many of the \\
    & foreign content   & because the tweets and sources provided  & foreign references.''\\
    & (2, 6) & were foreign.& \\
    
     \cline{2-4}
    & Heavy use of the & Whether the agent was dependable  & ``The fact that the source data were \\
    & same sources (1, 2) & in providing the same sources repeatedly. & from the same organization made the \\
    &  & & judgement seem a bit untrustworthy.'' \\

  \bottomrule
 \end{tabular}
 \end{table*}

The results support (H2) for (RQ2). Subjects relied on the agent's true consensus-based explanations significantly more than those based on false consensus. Significantly, \emph{the illusion of consensus did not appear}. This confirms the results of previous work \cite{connor_desai_getting_2022} for XAI explanations: by making the relationships among data sources transparent, our XAI agent prevented an illusion of consensus.
Still, unlike in previous work \cite{connor_desai_getting_2022, yousif_illusion_2019}, the difference between true and no consensus, i.e., the difference in number of sources used in the explanation, was not significantly different for reliance. The thematic analysis results suggest that subjects focused more on the trustworthiness of individual sources rather than on consensus and the number of data sources. Future work can explore how to raise awareness of this. 

This study was limited by its focus on trust and reliance over other influential factors, such as individual differences (e.g. neurotic tendencies), agent reliability, and so on. Incorporating these factors will help us better understand the impact of consensus-based explanations on trust and reliance. Some subjects were unable to use some data source tweet due to Twitter's volatility. Future studies should use research designs or technology hacks that prevent such events or explore their impact. As a lab-based study, our agent relied on a timeline containing 24 tweets. Larger, longer-term studies will be needed to better understand the dynamic nature of trust as well as studies in the wild. Finally, the agent could be explored as a general tool for, e.g., classifying toxicity or predicting risk of posting misinformation.





\section{Conclusion}
Source consensus in explanations affected reliance on the XAI agent, but trust was not a mediator. 
The illusion of consensus did not occur because the agent ensured that the relationships among the data sources were transparent.
Our findings provide initial evidence of the importance of revealing the relationships among data sources in explanations and the importance of providing true consensus in fact-checking XAI agents.



\begin{acks}
    This work was funded by a DLab Challenge: Laboratory for Design of Social Innovation in Global Networks (DLab) Research Grant.
    We thank Jacqueline Urakami and the Aspire Lab for early design and research feedback.
\end{acks}

\bibliographystyle{ACM-Reference-Format}
\bibliography{elemi-ks}

\appendix

\end{document}